\input{epsf}

\newcommand{\Set}{{\rm Set}}
\newcommand{\Hilb}{{\rm Hilb}}
\newcommand{\Cob}{{\rm Cob}}
\renewcommand{\to}{\rightarrow}
\newcommand{\maps}{\colon}

\documentstyle[12pt,amsfonts]{article}

\begin{document}

      \begin{center}
      {\bf Higher-Dimensional Algebra \\
      and Planck-Scale Physics  \\}
      \vspace{0.5cm}
      {\em John C.\ Baez\\}
      \vspace{0.3cm}
      {\small Department of Mathematics, University of California\\ 
      Riverside, California 92521 \\
      USA\\ }
      \vspace{0.3cm}
      {\small email: baez@math.ucr.edu\\}
      \vspace{0.3cm}
      {\small January 28, 1999 \\ }
      \vspace{0.3cm}
{\small To appear in {\sl Physics Meets Philosophy at the Planck Scale}, \\
eds.\ Craig Callender and Nick Huggett, Cambridge U.\ Press}
      \end{center}

\begin{abstract}

This is a nontechnical introduction to recent work on 
quantum gravity using ideas from higher-dimensional algebra.  
We argue that reconciling general relativity with the 
Standard Model requires a `background-free quantum theory with 
local degrees of freedom propagating causally'.  We describe
the insights provided by work on topological quantum field theories
such as quantum gravity in 3-dimensional spacetime.  These are
background-free quantum theories lacking local degrees of freedom,
so they only display some of the features we seek.  However, they
suggest a deep link between the concepts of `space' and `state',
and similarly those of `spacetime' and `process', which we argue 
is to be expected in any background-free quantum theory.  We sketch 
how higher-dimensional algebra provides the mathematical tools to
make this link precise.  Finally, we comment on attempts to formulate
a theory of quantum gravity in 4-dimensional spacetime using
`spin networks' and `spin foams'.   

\end{abstract}

\section{Introduction}

At present our physical worldview is deeply schizophrenic.  We have, not
one, but two fundamental theories of the physical universe: general
relativity, and the Standard Model of particle physics based on quantum
field theory.  The former takes gravity into account but ignores quantum
mechanics, while the latter takes quantum mechanics into account but
ignores gravity.  In other words, the former recognizes that spacetime
is curved but neglects the uncertainty principle, while the latter takes
the uncertainty principle into account but pretends that spacetime is
flat.  Both theories have been spectacularly successful in their own
domain, but neither can be anything more than an approximation to the
truth.   Clearly some synthesis is needed: at the very least, a theory 
of {\it quantum gravity}, which might or might not be part of a
overarching `theory of everything'.   Unfortunately, attempts to achieve
this synthesis have not yet succeeded.

Modern theoretical physics is difficult to understand for anyone outside
the subject.  Can philosophers really contribute to the project of
reconciling general relativity and quantum field theory?  Or is this a
technical business best left to the experts?  I would argue for the
former.  General relativity and quantum field theory are based on some
profound insights about the nature of reality.    These insights are
crystallized in the form of mathematics, but there is a limit to how
much progress we can make by just playing around with this mathematics.
We need to go back to the insights behind general relativity and quantum
field theory, learn to hold them together in our minds, and dare to
imagine a world more strange, more beautiful, but ultimately more {\it
reasonable} than our current theories of it.  For this daunting task,
philosophical reflection is bound to be of help.  

However, a word of warning is in order.  The paucity of experimental
evidence concerning quantum gravity has allowed research to proceed in a
rather unconstrained manner, leading to divergent schools of opinion. 
If one asks a string theorist about quantum gravity, one will get
utterly different answers than if one asks someone working on loop
quantum gravity or some other approach.  To make matters worse, experts
often fail to emphasize the difference between experimental results,
theories supported by experiment, speculative theories that have gained
a certain plausibility after years of study, and the latest fads. 
Philosophers must take what physicists say about quantum gravity with a
grain of salt.  

To lay my own cards on the table, I should say that as a mathematical
physicist with an interest in philosophy, I am drawn to a strand of work
that emphasizes `higher-dimensional algebra'.  This branch of
mathematics goes back and reconsiders some of the presuppositions that
mathematicians usually take for granted, such as the notion of equality
\cite{BD2} and the emphasis on doing mathematics using 1-dimensional
strings of symbols \cite{Brown,Kauffman}.  Starting in the late 1980s,
it became apparent that higher-dimensional algebra is the correct
language to formulate so-called `topological quantum field theories'
\cite{BD,L,Turaev}.  More recently, various people have begun to formulate
theories of quantum gravity using ideas from higher-dimensional algebra
\cite{B3,BC,FK,MS,Reis}.  While they have tantalizing connections to string
theory, these theories are best seen as an outgrowth of loop quantum
gravity \cite{RR}.  

The plan of the paper is as follows.  In Section \ref{planck.length}, I
begin by recalling why some physicists expect general relativity and
quantum field theory to collide at the Planck length.  This is a unit of
distance concocted from three fundamental constants: the speed of light
$c$, Newton's gravitational constant $G$, and Planck's constant $\hbar$.
General relativity idealizes reality by treating Planck's constant as
negligible, while quantum field theory idealizes it by treating Newton's
gravitational constant as negligible.  By analyzing the physics of
$c,G,$ and $\hbar$, we get a glimpse of the sort of theory that would be
needed to deal with situations where these idealizations break down.  In
particular, I shall argue that we need a {\it background-free quantum
theory with local degrees of freedom propagating causally}.  

In Section \ref{TQFT}, I discuss `topological quantum field theories'. 
These are the first examples of background-free quantum  theories. 
However, they lack local degrees of freedom.  In other words, they
describe imaginary worlds in which everywhere looks like everywhere
else!   This might at first seem to condemn them to the status of
mathematical curiosities.  However, they suggest an important analogy 
between the mathematics of spacetime and the mathematics of quantum
theory.   I argue that this is the beginning of a new bridge between
general relativity and quantum field theory.  

In Section \ref{3dQG}, I describe one of the most important examples of a
topological quantum field theory: the Turaev-Viro model of
quantum gravity in 3-dimensional spacetime.   This theory is just a 
warmup for the 4-dimensional case that is of real interest in physics.  
Nonetheless, it has some startling features which perhaps hint at the 
radical changes in our worldview that a successful synthesis of 
general relativity and quantum field theory would require.

In Section \ref{HDA}, I discuss the role of higher-dimensional algebra
in topological quantum field theory.  I begin with a brief
introduction to categories.  Category theory can be thought of as
an attempt to treat processes (or `morphisms') on an equal footing with
things (or `objects'), and it is ultimately for this reason that it
serves as a good framework for topological quantum field theory.  In
particular, category theory allows one to make the analogy between the
mathematics of spacetime and the mathematics of quantum theory quite
precise.  But to fully explore this analogy one must introduce
`$n$-categories', a generalization of categories that allows one to
speak of processes between processes between processes... and so on to
the $n$th degree.  Since $n$-categories are purely algebraic structures
but have a natural relationship to the study of $n$-dimensional
spacetime, their study is sometimes called `higher-dimensional algebra'.
 
Finally, in Section \ref{4dQG} I briefly touch upon recent attempts to
construct theories of 4-dimensional quantum gravity using 
higher-dimensional algebra.  This subjects is still in its infancy.
Throughout the paper, but especially  in this last section, the reader
must turn to the references for details.  To make the bibliography as
useful as possible, I have chosen references of an expository nature 
whenever they exist, rather than always citing the first paper in which 
something was done.  

\section{The Planck Length} \label{planck.length}

Two constants appear throughout general relativity: the speed of light
$c$ and Newton's gravitational constant $G$.   This should be no
surprise, since Einstein created general relativity to reconcile the
success of Newton's theory of gravity, based on instantaneous action at
a distance, with his new theory of special relativity, in which no
influence travels faster than light.  The constant $c$ also appears in
quantum field theory, but paired with a different partner: Planck's
constant $\hbar$.   The reason is that quantum field theory takes
into account special relativity and quantum theory, in which $\hbar$ 
sets the scale at which the uncertainty principle becomes important.  

It is reasonable to suspect that any theory reconciling general
relativity and quantum theory will involve all three constants $c$, $G$,
and $\hbar$.  Planck noted that apart from numerical factors there
is a unique way to use these constants to define units of length, time, 
and mass.  For example, we can define the unit of length now
called the `Planck length' as follows:
\[          \ell_P = \sqrt{\hbar G \over c^3}.  \]
This is extremely small: about $1.6 \cdot 10^{-35}$ meters.   Physicists
have long suspected that quantum gravity will become important for
understanding physics at about this scale.   The reason is very simple: any
calculation that predicts a length using only the constants $c$, $G$ and
$\hbar$ must give the Planck length, possibly multiplied by an
unimportant numerical factor like $2\pi$.  

For example, quantum field theory says that associated to any mass $m$
there is a length called its Compton wavelength, $\ell_C$, such that
determining the position of a particle of mass $m$ to within one Compton
wavelength requires enough energy to create another particle of that
mass.  Particle creation is a quintessentially quantum-field-theoretic
phenomenon.  Thus we may say that the Compton wavelength sets the
distance scale at which quantum field theory becomes crucial for
understanding the behavior of a particle of a given mass.  On the other
hand, general relativity says that associated to any mass $m$ there is a
length called the Schwarzschild radius, $\ell_S$, such that compressing
an object of mass $m$ to a size smaller than this results in the
formation of a black hole.  The Schwarzschild radius is roughly the
distance scale at which general relativity becomes crucial for
understanding the behavior of an object of a given mass.  Now, ignoring
some numerical factors, we have   
\[       \ell_C = {\hbar \over mc}  \]   
and  
\[       \ell_S = {Gm\over c^2}. \]   
These two lengths become equal when $m$ is the Planck mass.  And 
when this happens, they both equal the Planck length!

At least naively, we thus expect that {\it both} general relativity and
quantum field theory would be needed to understand the behavior of an
object whose mass is about the Planck mass and whose radius is about the
Planck length.  This not only explains some of the importance of the
Planck scale, but also some of the difficulties in obtaining
experimental evidence about physics at this scale.  Most of our
information about general relativity comes from observing heavy objects
like planets and stars, for which $\ell_S \gg \ell_C$.   Most of our
information about quantum field theory comes from observing light
objects like electrons and protons, for which $\ell_C \gg \ell_S$.  The
Planck mass is intermediate between these: about the mass of a largish
cell.  But the Planck length is about $10^{-20}$ times
the radius of a proton!   To study a situation where both general
relativity and quantum field theory are important, we could try to
compress a cell to a size $10^{-20}$ times that of a proton.   We
know no reason why this is impossible in principle.  But we have no idea
how to actually accomplish such a feat.  

There are some well-known loopholes in the above argument.   The
`unimportant numerical factor' I mentioned above might actually be very
large, or very small.  A theory of quantum gravity might make testable
predictions of dimensionless quantities like the ratio of the muon and
electron masses.  For that matter, a theory of quantum gravity might
involve physical constants other than $c$, $G$,  and $\hbar$.  The
latter two alternatives are especially plausible if we study quantum
gravity as part of a larger theory describing other forces and
particles.   However, even though we cannot prove that the Planck length
is significant for quantum gravity, I think we can glean some wisdom
from pondering the constants $c,G,$ and $\hbar$ --- and more
importantly, the physical insights that lead us to regard these
constants as important.

What is the importance of the constant $c$?   In special relativity,
what matters is the appearance of this constant in the Minkowski
metric 
\[                ds^2 = c^2 dt^2 - dx^2 - dy^2 - dz^2  \] 
which defines the geometry of spacetime, and in particular the lightcone
through each point.  Stepping back from the specific formalism here, we
can see several ideas at work.  First, space and time form a unified
whole which can be thought of geometrically.  Second, the quantities
whose values we seek to predict are localized.  That is, we can measure
them in small regions of spacetime (sometimes idealized as points).  
Physicists call such quantities `local degrees of freedom'.  And third,
to predict the value of a quantity that can be measured in some region
$R$, we only need to use values of quantities measured in regions that
stand in a certain geometrical relation to $R$.  This relation is called
the `causal structure' of spacetime.  For example, in a relativistic
field theory, to predict the value of the fields in some region $R$, it
suffices to use their values in any other region that intersects every
timelike path passing through $R$.  The common way of summarizing this
idea is to say that nothing travels faster than light.  I prefer to say
that a good theory of physics should have {\it local degrees of freedom
propagating causally}. 

In Newtonian gravity, $G$ is simply the strength of the gravitational
field.  It takes on a deeper significance in general relativity, where
the gravitational field is described in terms of the curvature of the
spacetime metric.  Unlike in special relativity, where the Minkowski 
metric is a `background structure' given a priori, in general relativity
the metric is treated as a field which not only affects, but also is
affected by, the other fields present.  In other words, the geometry of
spacetime becomes a local degree of freedom of the theory.
Quantitatively, the interaction of the metric and other fields is
described by Einstein's equation
\[                 G_{\mu \nu} = 8\pi G T_{\mu \nu} ,\]
where the Einstein tensor $G_{\mu \nu}$ depends on the curvature of the
metric, while the stress-energy tensor $T_{\mu \nu}$ describes the flow
of energy and momentum due to all the other fields.  The role of the
constant $G$ is thus simply to quantify how much the geometry of
spacetime is affected by other fields.   Over the years, people have
realized that the great lesson of general relativity is that a good
theory of physics should contain no geometrical structures that affect
local degrees of freedom while remaining unaffected by them.  Instead,
all geometrical structures --- and in particular the causal structure ---
should themselves be local degrees of freedom.  For short, one says
that the theory should be {\it background-free}.
 
The struggle to free ourselves from background structures began long
before Einstein developed general relativity, and is still not complete.
The conflict between Ptolemaic and Copernican cosmologies, the dispute
between Newton and Leibniz concerning absolute and relative motion, and
the modern arguments concerning the `problem of time' in quantum gravity
--- all are but chapters in the story of this struggle.  I do not have
room to sketch this story here, nor even to make more precise the
all-important notion of `geometrical structure'.  I can only point the
reader towards the literature, starting perhaps with the books by
Barbour \cite{Barbour} and Earman \cite{Earman}, various papers by
Rovelli \cite{Rovelli1,Rovelli2,Rovelli3}, and the many references therein.  

Finally, what of $\hbar$?  In quantum theory, this appears most 
prominently in the commutation relation between the momentum $p$ and
position $q$ of a particle: 
\[                 pq - qp = -i\hbar  ,\]
together with similar commutation relations involving other pairs of
measurable quantities.   Because our ability to measure two quantities
simultaneously with complete precision is limited by their failure to
commute, $\hbar$ quantifies our inability to simultaneously know
everything one might choose to know about the world.  But there is far
more to quantum theory than the uncertainty principle.  In practice,
$\hbar$ comes along with the whole formalism of complex Hilbert spaces
and linear operators.  

There is a widespread sense that the principles behind quantum theory
are poorly understood compared to those of general relativity.  This has
led to many discussions about interpretational issues.  However, I do
not think that quantum theory will lose its mystery through such
discussions.   I believe the real challenge is to better understand why
the mathematical formalism of quantum theory is precisely what it is. 
Research in quantum logic has done a wonderful job of understanding the
field of candidates from which the particular formalism we use has been
chosen.  But what is so special about this particular choice? 
Why, for example, do we use complex Hilbert spaces rather than real or
quaternionic ones?  Is this decision made solely to fit the experimental
data, or is there a deeper reason?  Since questions like this do not yet
have clear answers, I shall summarize the physical insight behind
$\hbar$ by saying simply that a good theory of the physical universe
should be a {\it quantum theory} --- leaving open the possibility of
eventually saying something more illuminating.

Having attempted to extract the ideas lying behind the constants $c,G,$
and $\hbar$, we are in a better position to understand the task of
constructing a theory of quantum gravity.  General relativity
acknowledges the importance of $c$ and $G$ but idealizes reality by
treating $\hbar$ as negligibly small.  From our discussion above, we see
that this is because general relativity is a background-free
classical theory with local degrees of freedom propagating causally.
On the other hand, quantum field theory as normally practiced
acknowledges $c$ and $\hbar$ but treats $G$ as negligible, because it is
a background-dependent quantum theory with local degrees of freedom
propagating causally.

The most conservative approach to quantum gravity is to seek a theory
that combines the best features of general relativity and quantum field
theory.    To do this, we must try to find a {\it background-free
quantum theory with local degrees of freedom propagating causally}.  
While this approach may not succeed, it is definitely worth pursuing.
Given the lack of experimental evidence that would point us towards 
fundamentally new principles, we should do our best to understand
the full implications of the principles we already have!

From my description of the goal one can perhaps see some of the
difficulties.  Since quantum gravity should be background-free, the
geometrical structures defining the causal structure of spacetime should
themselves be local degrees of freedom propagating causally.  This much
is already true in general relativity.  But because quantum gravity
should be a quantum theory, these degrees of freedom should be treated
quantum-mechanically.  So at the very least, we should develop a quantum
theory of some sort of geometrical structure that can define a causal
structure on spacetime.   

String theory has not gone far in this direction.  This  theory is
usually formulated with the help of a metric on spacetime, which is
treated as a background structure rather than a local degree of freedom
like the rest.   Most string theorists recognize that this is an
unsatisfactory situation, and by now many are struggling towards a
background-free formulation of the theory.   However, in the words of
two experts \cite{HN}, ``it seems that a still more radical departure
from conventional ideas about space and time may be required in order to
arrive at a truly background independent formulation.''

Loop quantum gravity has gone a long way towards developing a
background-free quantum theory of the geometry of space
\cite{Ashtekar,Rovelli4}, but not so far when it comes to spacetime. 
This has made it difficult to understand dynamics, and particular the
causal propagation of degrees of freedom.  Work in earnest on these
issues has begun only recently.  One reason for optimism is the recent
success in understanding quantum gravity in 3 spacetime dimensions.  
But to explain this, I must first say a bit about topological quantum
field theory.  

\section{Topological Quantum Field Theory} \label{TQFT}

Besides general relativity and quantum field theory as usually
practiced, a third sort of idealization of the physical world has
attracted a great deal of attention in the last decade.  These are
called topological quantum field theories, or `TQFTs'.  In the
terminology of the previous section, a TQFT is a {\it background-free
quantum theory with no local degrees of freedom}\footnote{It would be
nicely symmetrical if TQFTs involved the constants $G$ and $\hbar$ but
not $c$.  Unfortunately I cannot quite see how to make this idea precise.}.

A good example is quantum gravity in 3-dimensional spacetime.  First let
us recall some features of {\it classical} gravity in 3-dimensional
spacetime.  Classically, Einstein's equations predict qualitatively very
different phenomena depending on the dimension of spacetime.  If
spacetime has 4 or more dimensions, Einstein's equations imply that the
metric has local degrees of freedom.   In other words, the curvature of
spacetime at a given point is not completely determined by the flow of
energy and momentum through that point: it is an independent variable in
its own right.  For example, even in the vacuum, where the
energy-momentum tensor vanishes, localized ripples of curvature can
propagate in the form of gravitational radiation.  In 3-dimensional
spacetime, however, Einstein's equations suffice to completely determine
the curvature at a given point of spacetime in terms of the flow of
energy and momentum through that point.    We thus say that the metric
has no local degrees of freedom.   In particular, in the vacuum the
metric is flat, so every small patch of empty spacetime looks
exactly like every other.  

The absence of local degrees of freedom makes general relativity far
simpler in 3-dimensional spacetime than in higher dimensions. Perhaps
surprisingly, it is still somewhat interesting.  The reason is the
presence of `global' degrees of freedom.   For example, if we chop a
cube out of flat 3-dimensional Minkowski space and form a 3-dimensional
torus by identifying the opposite faces of this cube, we get a spacetime
with a flat metric on it, and thus a solution of the vacuum Einstein
equations.  If we do the same starting with a larger cube, or a
parallelipiped, we get a different spacetime that also satisfies the
vacuum Einstein equations.  The two spacetimes are {\it locally}
indistinguishable, since locally both look just like flat Minkowski
spacetime.  However, they can be distinguished {\it globally} ---  for
example, by measuring the volume of the whole spacetime, or studying
the behavior of geodesics that wrap all the way around the torus.  

Since the metric has no local degrees of freedom in 3-dimensional 
general relativity, this theory is much easier to quantize than the
physically relevant 4-dimensional case.   In the simplest situation,
where we consider `pure' gravity without matter, we obtain a
background-free quantum field theory with no local degrees of freedom
whatsoever: a TQFT.  

I shall say more about 3-dimensional quantum gravity in Section
\ref{3dQG}.  To set the stage, let me sketch the axiomatic approach to 
topological quantum field theory proposed by Atiyah \cite{Atiyah}.  My
earlier definition of a TQFT as a `background-free quantum field theory
with no local degrees of freedom' corresponds fairly well to how
physicists think about TQFTs.   But mathematicians who wish to prove
theorems about TQFTs need to start with something more precise, so
they often use Atiyah's axioms.

An important feature of TQFTs is that they do not presume a fixed
topology for space or spacetime.  In other words, when dealing with an
$n$-dimensional TQFT, we are free to choose any $(n-1)$-dimensional
manifold to represent space at a given time\footnote{Here and in  what
follows, by `manifold' I really mean `compact oriented smooth manifold',
and cobordisms between these will also be compact, oriented, and
smooth.}.  Moreover given two such manifolds, say $S$ and $S'$, we are
free to choose any $n$-dimensional  manifold $M$ to represent the
portion of spacetime between $S$ and $S'$. Mathematicians call $M$ a
`cobordism' from $S$ to $S'$.  We write $M \maps S \to S'$, because we
may think of $M$ as the process of time passing from the moment $S$ to
the moment $S'$.  

\begin{figure}[ht]
\vskip 2em
\centerline{\epsfysize=1.5in\epsfbox{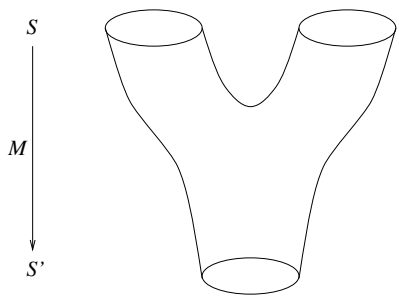}} \medskip
\caption{A cobordism}
\label{cobordism}
\end{figure}
 
For example, in Figure \ref{cobordism} we depict a 2-dimensional
manifold $M$ going from a 1-dimensional manifold $S$ (a pair of circles)
to a 1-dimensional manifold  $S'$ (a single circle).   Crudely speaking,
$M$ represents a process in which two separate spaces collide to form a
single one!  This may seem outr\'e, but these days physicists are
quite willing to speculate about processes in which the topology of
space changes with the passage of time.  Other forms of topology change
include the the formation of a wormhole, the appearance of the universe
in a `big bang', or its disappearance in a `big crunch'.  

\begin{figure}[h]
\vskip 2em
\centerline{\epsfysize=2in\epsfbox{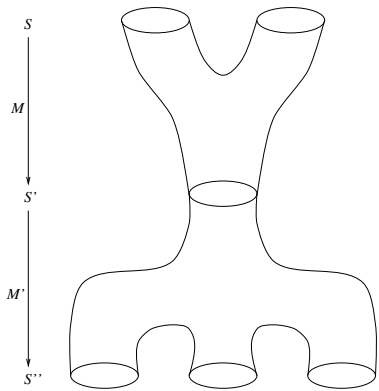}} \medskip
\caption{Composition of cobordisms}
\label{composition}
\end{figure}

There are various important operations one can perform on cobordisms,
but I will only describe two.  First, we may `compose' two cobordisms 
$M\maps S \to S'$ and $M \maps S' \to S''$, obtaining a cobordism 
$M' M \maps S \to S''$, as
illustrated in Figure \ref{composition}.  The idea here is that the
passage of time corresponding to $M$ followed by the passage of time
corresponding to $M'$ equals the passage of time corresponding to 
$M'M$.   This is analogous to the familiar idea that waiting $t$ seconds
followed by waiting $t'$ seconds is the same as waiting $t+t'$ seconds. 
The big difference is that in topological quantum field theory we cannot
measure time in seconds, because there is no background metric available
to let us count the passage of time!  We can only keep track of topology
change.  Just as ordinary addition is associative, composition of
cobordisms satisfies the associative law:
\[                (M'' M') M = M'' (M' M)     .        \] 
However, composition of cobordisms is not commutative.  As we shall
see, this is related to the famous noncommutativity of observables
in quantum theory.

\begin{figure}[h]
\vskip 2em
\centerline{\epsfysize=1.5in\epsfbox{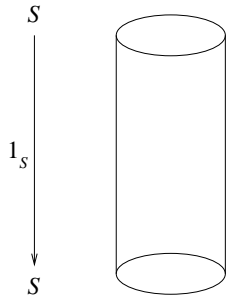}} \medskip
\caption{An identity cobordism}
\label{identity}
\end{figure}

Second, for any $(n-1)$-dimensional manifold $S$ representing space,
there is a cobordism $1_S \maps S \to S$ called the `identity'
cobordism, which represents a passage of time without any topology
change.  For example, when $S$ is a circle, the identity cobordism
$1_S$ is a cylinder, as shown in Figure \ref{identity}.   In general,
the identity cobordism $1_S$ has the property that for any cobordism
$M \maps S' \to S$ we have
\[                         1_S M = M , \]
while for any cobordism $M \maps S \to S'$ we have
\[                         M 1_S  = M .\]
These properties say that an identity cobordism is analogous to waiting 0
seconds: if you wait 0 seconds and then wait $t$ more seconds, or wait
$t$ seconds and then wait 0 more seconds, this is the same as waiting
$t$ seconds.

These operations just formalize of the notion of `the passage of
time' in a context where the topology of spacetime is arbitrary and
there is no background metric.    Atiyah's axioms relate this notion to
quantum theory as follows.   First, a TQFT must assign a Hilbert space
$Z(S)$ to each $(n-1)$-dimensional manifold $S$.  Vectors in this
Hilbert space represent possible states of the universe given that space
is the  manifold $S$.  Second, the TQFT must assign a linear operator
$Z(M) \maps Z(S) \to Z(S')$ to each $n$-dimensional cobordism $M \maps S
\to S'$.  This operator describes how states change given that the
portion of spacetime between $S$ and $S'$ is the manifold $M$. In other
words, if space is initially the manifold $S$ and the state of  the
universe is $\psi$, after the passage of time corresponding to $M$ the
state of the universe will be $Z(M)\psi$.  

In addition, the TQFT must satisfy a list of properties.  Let me just
mention two.  First, the TQFT must preserve composition.  That is,
given cobordisms $M \maps S \to S'$ and $M' \maps S' \to S''$, we must
have 
\[                Z(M'M) = Z(M')Z(M),  \] 
where the right-hand side denotes the composite of the operators
$Z(M)$ and $Z(M')$.  Second, it must preserve identities.  That is, given
 any manifold $S$ representing space, we must have 
\[                Z(1_S) = 1_{Z(S)}  .\] 
where the right-hand side denotes the identity operator on the
Hilbert space $Z(S)$.   

Both these axioms are eminently reasonable if one ponders them a bit. 
The first says that the passage of time corresponding to the cobordism 
$M$ followed by the passage of time corresponding to $M'$ has the  same
effect on a state as the combined passage of time corresponding to
$M'M$.  The second says that a passage of time in which no topology 
change occurs has no effect at all on the state of the universe.  This
seems paradoxical at first, since it seems we regularly observe things 
happening even in the absence of topology change.   However, this
paradox is easily resolved: a TQFT describes a world quite unlike ours,
one without local degrees of freedom.  In such a world, nothing local
happens, so the state of the universe can only change when the topology
of space itself changes\footnote{Actually, while perfectly correct as
far as it goes, this resolution dodges an important issue.  Some
physicists have suggested that the second axiom may hold even in quantum
field theories {\it with} local degrees of freedom, so long as they are
background-free \cite{Barrett}. Unfortunately a discussion of this would
take us too far afield here.}.

The most interesting thing about the TQFT axioms is their common formal 
character.    Loosely speaking, they all say that a TQFT maps structures
in differential topology --- by which I mean the study of manifolds ---
to corresponding structures in quantum theory.  In coming up with these
axioms, Atiyah took advantage of a powerful analogy between differential
topology and quantum theory, summarized in Table 1.  

\vskip 2em
\begin{center}
{\small
\begin{tabular}{|c|c|}                    \hline
DIFFERENTIAL TOPOLOGY             &  QUANTUM THEORY             \\  \hline
$(n-1)$-dimensional manifold      &  Hilbert space              \\  
(space)                           &  (states)                   \\  \hline
cobordism between $(n-1)$-dimensional manifolds &  operator     \\ 
(spacetime)                       &  (process)                  \\  \hline
composition of cobordisms         &  composition of operators   \\  \hline
identity cobordism                &  identity operator          \\  \hline
\end{tabular}} \vskip 1em
Table 1.  Analogy between differential topology and quantum theory
\end{center}
\vskip 0.5em

I shall explain this analogy between differential topology and quantum
theory further in Section \ref{HDA}.   For now, let me just emphasize
that this analogy is exactly the sort of clue we should pursue for a
deeper understanding of quantum gravity.   At first glance, general
relativity and quantum theory look very different mathematically: one
deals with space and spacetime, the other with Hilbert spaces and
operators.  Combining them has always seemed a bit like mixing oil and
water.  But topological quantum field theory suggests that perhaps they
are not so different after all!  Even better, it suggests a concrete
program of synthesizing the two, which many mathematical physicists are
currently pursuing. Sometimes this goes by the name of `quantum
topology' \cite{BK,Turaev}.

Quantum topology is very technical, as anything involving mathematical
physicists inevitably becomes.   But if we stand back a moment, it
should be perfectly obvious that differential topology and quantum
theory must merge if we are to understand background-free quantum field
theories.  In physics that ignores general relativity, we treat space as
a background on which states of the world are displayed.  Similarly, we
treat spacetime as a background on which the process of change occurs. 
But these are idealizations which we must overcome in a background-free
theory.   In fact, the concepts of `space' and `state' are two aspects
of a unified whole, and likewise for the concepts of `spacetime' and
`process'.   It is a challenge, not just for mathematical physicists,
but also for philosophers, to understand this more deeply.

\section{3-Dimensional Quantum Gravity} \label{3dQG}

Before the late 1980s, quantum gravity was widely thought to be just as
intractable in 3 spacetime dimensions as in the physically important
4-dimensional case.  The situation changed drastically when physicists
and mathematicians developed the tools for handling background-free
quantum theories without local degrees of freedom.  By now, it is easier
to give a complete description of 3-dimensional quantum gravity than
most quantum field theories of the traditional sort!   

Let me sketch how one sets up a theory of 3-dimensional quantum gravity
satisfying Atiyah's axioms for a TQFT.    Before doing so I should warn
reader that there are a number of inequivalent theories of 3-dimensional
quantum gravity \cite{Carlip}.  The one I shall describe is called the
Turaev-Viro model \cite{Turaev}.   While in some ways this is not the
most physically realistic one, since it is a quantum theory of Riemannian
rather than Lorentzian metrics, it illustrates the points I want to make
here.  

To get a TQFT satisfying Atiyah's axioms we need to describe a Hilbert
space of states for each 2-dimensional manifold and an operator for each
cobordism between 2-dimensional manifolds.  We begin by constructing a
preliminary Hilbert space $\tilde{Z}(S)$ for any 2-dimensional manifold 
$S$.  This construction requires choosing a background structure: a way
of chopping $S$ into triangles.   Later we will eliminate this
background-dependence and construct the Hilbert space of real physical
interest.   

\begin{figure}[h]
\vskip 2em
\centerline{\epsfysize=2in\epsfbox{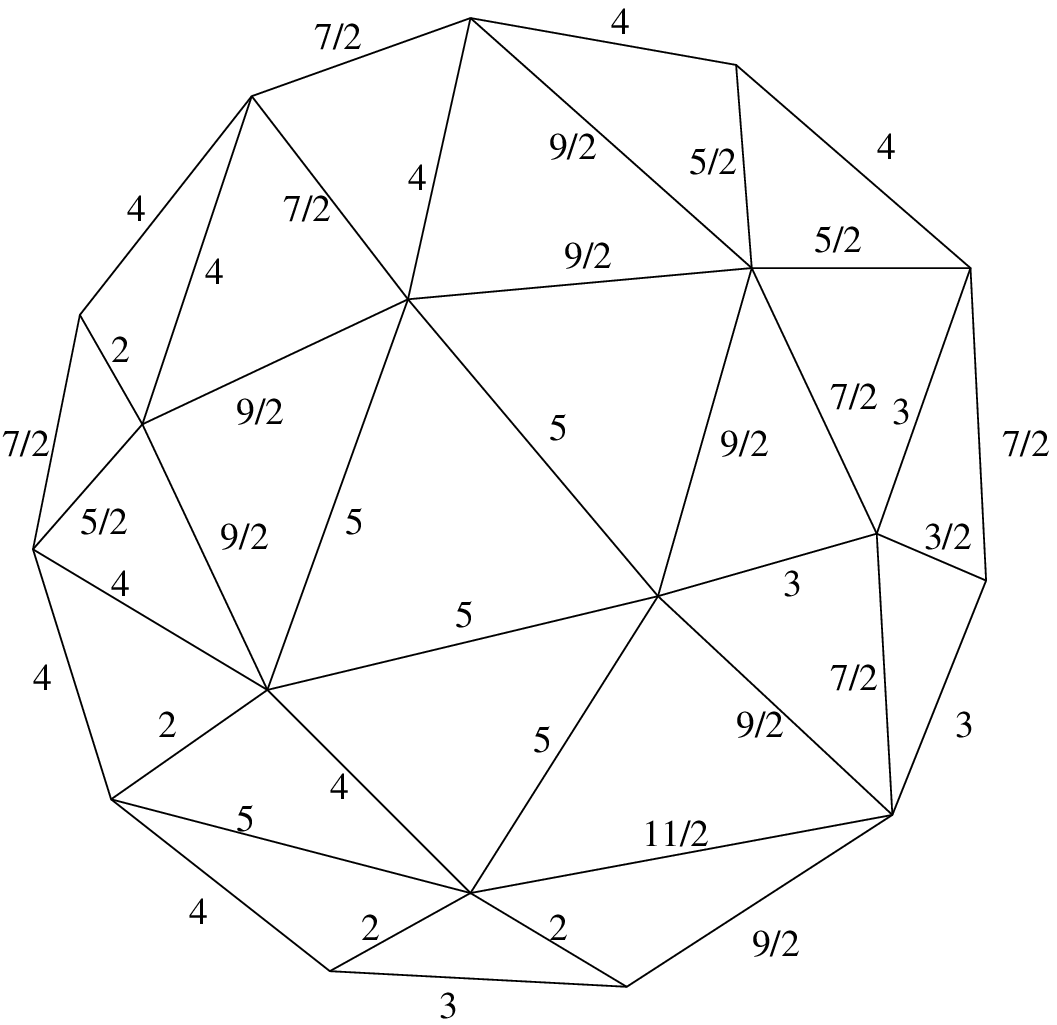}} \medskip
\caption{A state in the preliminary Hilbert space 
for 3-dimensional quantum gravity}
\label{triangulated-surface}
\end{figure}

To define the Hilbert space $\tilde{Z}(S)$, it is enough to specify an
orthonormal basis for it.  We decree that states in this basis are ways
of labelling the edges of the triangles in $S$ by numbers of the form
$0,{1\over 2},1,{3\over 2},\dots,{k\over 2}$.  An example is shown in 
Figure \ref{triangulated-surface}, where we take $S$ to be a sphere.   
Physicists call the numbers labelling the edges `spins',
alluding to the fact that we are using mathematics developed in the
study of angular momentum.  But here these numbers represent the {\it
lengths} of the edges as measured in units of the Planck length.  In
this theory, length is a discrete rather than continuous quantity!   

Then we construct an operator $\tilde{Z}(M) \maps \tilde{Z}(S) \to
\tilde{Z}(S')$ for each cobordism $M \maps S \to S'$.  Again we do this
with the help of a background structure on $M$: we choose a way to chop
it into tetrahedra, whose triangular faces must include among them the
triangles of $S$ and $S'$.  To define $\tilde{Z}(M)$ it is enough to
specify the transition amplitudes $\langle\psi',\tilde{Z}(M)\psi\rangle$
when $\psi$ and $\psi'$ are states in the bases given above.    We do
this as follows.  The states $\psi$ and $\psi'$ tell us how to label the
edges of triangles in $S$ and $S'$ by spins.  Consider any way to label
the edges of $M$ by spins that is compatible with these labellings of
edges in $S$ and $S'$.  We can think of this as a `quantum geometry' for
spacetime, since it tells us the shape of every tetrahedron in $M$. 
Using a certain recipe we can compute a complex number for this
geometry, which we think of as its `amplitude' in the quantum-mechanical
sense.  We then sum these amplitudes over all geometries to get the
total transition amplitude from $\psi$ to $\psi'$.  The reader familiar
with quantum field theory may note that this  construction is a discrete
version of a `path integral'.  

Now let me describe how we erase the background-dependence from this
construction.    Given an identity cobordism $1_S \maps S \to S$, the
operator $\tilde{Z}(1_S)$ is usually not the identity, thus violating
one of Atiyah's axioms for a topological quantum field theory.  However,
the next best thing happens: this operator maps $\tilde{Z}(S)$ onto a
subspace, and it acts as the identity on this subspace.  This subspace,
which we call $Z(S)$, is the Hilbert space of real physical interest in
3-dimensional quantum gravity.  Amazingly, this subspace doesn't depend
on how we chopped $S$ into triangles.  Even better, for any cobordism $M
\maps S \to S'$, the operator $\tilde{Z}(M)$ maps $Z(S)$ to $Z(S')$.  
Thus it restricts to an operator $Z(M) \maps Z(S) \to Z(S')$.  Moreover,
this operator $Z(M)$ turns out not to depend on how we chopped $M$ into
tetrahedra.   To top it all off, it turns out that the Hilbert spaces
$Z(S)$ and operators $Z(M)$ satisfy Atiyah's axioms.

In short, we started by chopping space into triangles and spacetime into
tetrahedra, but at the end of the day nothing depends on this choice of
background structure.  It also turns out that the final theory has no
local degrees of freedom: all the measurable quantities are global in
character.   For example, there is no operator on $Z(S)$ corresponding
to the `length of a triangle's edge', but there is an operator
corresponding to the length of the shortest geodesic wrapping around the
space $S$ in a particular way.  These miracles are among the main
reasons for interest in quantum topology.  They only happen because of
the carefully chosen recipe for computing amplitudes for spacetime
geometries.   This recipe is the real core of the whole construction. 
Sadly, it is a bit too technical to describe here, so the reader will
have to turn elsewhere for details \cite{Kauffman,Turaev}.  I can say
this, though: the reason this recipe works so well is that it neatly
combines ideas from general relativity, quantum field theory, and a
third subject that might at first seem unrelated --- higher-dimensional
algebra.

\section{Higher-Dimensional Algebra} \label{HDA}

One of the most remarkable accomplishments of the early 20th century 
was to formalize all of mathematics in terms of a language with a
deliberately impoverished vocabulary: the language of set theory.  In
Zermelo-Fraenkel set theory, everything is a set, the only fundamental
relationships between sets are membership and equality, and two sets are
equal if and only if they have the same elements.  If in Zermelo-Fraenkel
set theory you ask what sort of thing is the number $\pi$, the
relationship `less than', or the exponential function, the answer is
always the same: a set!  Of course one must bend over backwards to think
of such varied entities as sets, so this formalization may seem almost
deliberately perverse.   However, it represents the culmination of a
worldview in which things are regarded as more fundamental than processes 
or relationships.

More recently, mathematicians have developed a somewhat more flexible
language, the language of category theory.  Category theory is an attempt
to put processes and relationships on an equal status with things.  A
category consists of a collection of `objects', and for each pair of
objects $x$ and $y$, a collection of `morphisms' from $x$ to $y$.  We
write a morphism from $x$ to $y$ as $f \maps x \to y$.   We demand
that for any morphisms $f \maps x \to y$ and $g \maps y \to z$, we can
`compose' them to obtain a morphism $gf \maps x \to z$.  We also demand
that composition be associative.  Finally, we demand that for any object
$x$ there be a morphism $1_x$, called the `identity' of $x$, such that 
$f1_x = f$ for any morphism $f \maps x \to y$ and $1_x g = g$ for any 
morphism $g \maps y \to x$. 

Perhaps the most familiar example of a category is $\Set$.  Here the
objects are sets and the morphisms are functions between sets.  
However, there are many other examples.  Fundamental to quantum theory
is the category $\Hilb$.  Here the objects are complex Hilbert spaces
and the morphisms are linear operators between Hilbert spaces.  In
Section \ref{TQFT} we also met a category important in differential
topology, the category $n\Cob$.  Here the objects are
$(n-1)$-dimensional manifolds and the morphisms are cobordisms between
such manifolds.  Note that in this example, the morphisms are not
functions!  Nonetheless we can still think of them as `processes' going
from one object to another.  

An important part of learning category theory is breaking certain habits
one may have acquired from set theory.  For example, in category theory
one must resist the temptation to `peek into the objects'.   Traditionally, 
the first thing one asks about a set is: what are its elements?  A set
is like a container, and the contents of this container are the most
interesting thing about it.   But in category theory, an object need not
have `elements' or any sort of internal structure.  Even if it does,
this is not what really matters!   What really matters about an object
is its morphisms to and from other objects.   Thus category theory
encourages a relational worldview in which things are described, not in
terms of their constituents, but by their relationships to other things.
 
Category theory also downplays the importance of equality between
objects.  Given two elements of a set, the first thing one asks about
them is: are they equal?  But for objects in a category, we should ask
instead whether they are isomorphic.  Technically, the objects $x$ and
$y$ are said to be `isomorphic' if there is an morphism $f \maps x \to
y$ that has an `inverse': a morphism $f^{-1} \maps y \to x$ for which
$f^{-1}f = 1_x$ and $ff^{-1} = 1_y$.  A morphism with an inverse is
called an `isomorphism'.  An isomorphism between two objects lets turn
any morphism to or from one of them into a morphism to or from the other
in a reversible sort of way.  Since what matters about objects are their
morphisms to and from other objects, specifying an isomorphism between
two objects lets us treat them as `the same' for all practical purposes.

\begin{figure}[h]
\vskip 2em
\centerline{\epsfysize=1.2in\epsfbox{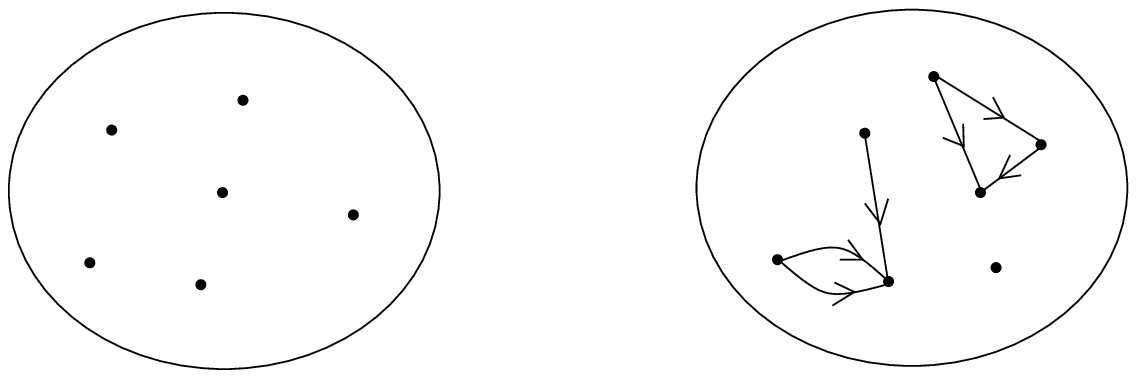}} \medskip
\caption{A set and a category}
\label{categories}
\end{figure}

Categories can be regarded as higher-dimensional analogs of sets.  As
shown in Fig.\ \ref{categories}, we may visualize a set as a bunch of
points, namely its elements.  Similarly, we may visualize a category as
a bunch of points corresponding to its objects, together with a bunch of
1-dimensional arrows corresponding to its morphisms.  (For simplicity, I
have not drawn the identity morphisms in Fig.\ 5.)

We may use the analogy between sets and categories to `categorify'
almost any set-theoretic concept, obtaining a category-theoretic
counterpart \cite{BD2}.  For example, just as there are functions
between sets, there are `functors' between categories.   A function from
one set to another sends each element of the first to an element of the
second.  Similarly, a functor $F$ from one category to another sends
each object $x$ of the first to an object $F(x)$ of the second, and
also sends each morphism $f \maps x \to y$ of the first to a morphism
$F(f) \maps F(x) \to F(y)$ of the second.  In addition, functors are
required to preserve composition and identities:   
\[          F(f' f) = F(f') F(f) \]   
and    
\[           F(1_x) = 1_{F(x)} .\]
Functors are important because they allow us to apply the relational 
worldview discussed above, not just to objects in a given category, but
to categories themselves.  Ultimately what matters about a category is
not its `contents' --- its objects and morphisms --- but its functors to
and from other categories!
 
\vskip 2em
\begin{center}
{\small
\begin{tabular}{|c|c|}                    \hline
SET THEORY  &  CATEGORY THEORY               \\     \hline
elements    &  objects                       \\     \hline
equations between elements  & isomorphisms between objects    \\     \hline
sets        &  categories                    \\     \hline
functions between sets  &  functors between categories     \\  \hline
equations between functions & natural isomorphisms between functors \\  \hline
\end{tabular}} \vskip 1em
Table 2.  Analogy between set theory and category theory
\end{center}
\vskip 0.5em

We summarize the analogy between set theory and category theory in 
Table 2.  In addition to the terms already discussed there is a
concept of `natural isomorphism' between functors.  This is the
correct analog of an equation between functions, but we will not need
it here --- I include it just for the sake of completeness.  

The full impact of category-theoretic thinking has taken a while to be
felt.  Categories were invented in the 1940s by Eilenberg and Mac Lane
for the purpose of clarifying relationships between algebra and
topology.  As time passed they became increasingly recognized as a
powerful tool for exploiting analogies throughout mathematics
\cite{MacLane}.   In the early 1960s they led to revolutionary --- and
still controversial --- developments in mathematical logic
\cite{Goldblatt}.  It gradually became clear that category theory
was a part of a deeper subject, `higher-dimensional algebra', in which
the concept of a category is generalized to that of an `$n$-category'.
But only by the 1990s did the real importance of categories for physics
become evident, with the discovery that higher-dimensional algebra is
the perfect language for topological quantum field theory \cite{Crane,L}.

Why are categories important in topological quantum field theory? 
The most obvious answer is that a TQFT is a functor.  Recall from Section
\ref{TQFT} that a TQFT maps each manifold $S$ representing space to a
Hilbert space $Z(S)$ and each cobordism $M \maps S \to S'$ representing
spacetime to an operator $Z(M) \maps Z(S) \to Z(S')$, in such a
way that composition and identities are preserved.  We may summarize all
this by saying that a TQFT is a functor 
\[   Z \maps   n\Cob \to \Hilb.  \]
In short, category theory makes the analogy in Table 1 completely
precise.  In terms of this analogy, many somewhat mysterious aspects of
quantum theory correspond to easily understood facts about spacetime!
For example, the noncommutativity of operators in quantum
theory corresponds to the noncommutativity of composing cobordisms. 
Similarly, the all-important `adjoint' operation in quantum theory,
which turns an operator $A \maps  H \to H'$ into an operator $A^* \maps
H' \to H$, corresponds to the operation of reversing the roles of past
and future in a cobordism $M \maps S \to S'$, obtaining a cobordism $M^*
\maps S' \to S$.  

But the role of category theory goes far beyond this.  The real surprise
comes when one examines the details of specific TQFTs.  In Section
\ref{3dQG} I sketched the construction of 3-dimensional quantum gravity,
but I left out the recipe for computing amplitudes for spacetime
geometries.   Thus the most interesting features of the whole business
were left as unexplained `miracles': the background-independence of the
Hilbert spaces $Z(S)$ and operators $Z(M)$, and the fact that they
satisfy Atiyah's axioms for a TQFT.   In fact, the recipe for amplitudes
and the verification of these facts make heavy use of category theory. 
The same is true for all other theories for which Atiyah's axioms have
been verified.  For some strange reason, it seems that category theory is
precisely suited to explaining what makes a TQFT tick.

For the last 10 years or so, various researchers have been trying to
understand this more deeply.  Much remains mysterious, but it now seems
that TQFTs are intimately related to category theory because of special
properties of the category $n\Cob$.  While $n\Cob$ is defined using
concepts from differential topology, a great deal of evidence suggests
that it admits a simple description in terms of `$n$-categories'.

I have already alluded to the concept of `categorification' --- the
process of replacing sets by categories, functions by functors and so
on, as indicated in Table 2.   The concept of `$n$-category' is obtained
from the concept of `set' by categorifying it $n$ times!   An
$n$-category has objects, morphisms between objects, 2-morphisms between
morphisms, and so on up to $n$-morphisms, together with various
composition operations satisfying various reasonable laws \cite{B2}.  
Increasing the value of $n$ allows an ever more nuanced treatment of
the notion of `sameness'.  A 0-category is just a set, and in a set 
the elements are simply equal or unequal.  A 1-category is a category, 
and in this context we may speak not only of equal but also of isomorphic 
objects.  Unfortunately, this careful distinction between equality and 
isomorphism breaks down when we study the morphisms.  Morphisms in a 
category are either the same or different; there is no concept of 
isomorphic morphisms.  In a 2-category this is remedied by introducing 
2-morphisms between morphisms.  Unfortunately, in a 2-category we cannot 
speak of isomorphic 2-morphisms.  To remedy this we must introduce the 
notion of 3-category, and so on.

We may visualize the objects of an $n$-category as points, the morphisms
as arrows going between these points, the 2-morphisms as 2-dimensional
surfaces going between these arrows, and so on.  There is thus a natural
link between $n$-categories and $n$-dimensional topology.  Indeed, one
reason why $n$-categories are a bit formidable is that calculations with
them are most naturally done using $n$-dimensional diagrams.   But this
link between $n$-categories and $n$-dimensional topology is precisely
why there may be a nice description of $n\Cob$ in the language of
$n$-categories.   

Dolan and I have proposed such a description, which we call the
`cobordism hypothesis' \cite{BD}.  Much work remains to be done to
make this hypothesis precise and prove or disprove it.  Proving it would
lay the groundwork for understanding topological quantum field theories
in a systematic way.  But beyond this, it would help us towards a
purely algebraic understanding of `space' and `spacetime' --- which is
precisely what we need to marry them to the quantum-mechanical notions of
`state' and `process'.  

\section{4-Dimensional Quantum Gravity}  \label{4dQG}

How important are the lessons of topological quantum field theory for
4-dimensional quantum gravity?  This is still an open question.  Since 
TQFTs lack local degrees of freedom, they are at best a warmup for the
problem we really want to tackle: constructing a background-free quantum
theory with local degrees of freedom propagating causally.  Thus, even
though work on TQFTs has suggested new ideas linking quantum theory and 
general relativity, these ideas may be too simplistic to be useful in
real-world physics.

However, physics is not done by sitting on ones hands and pessimistically
pondering the immense magnitude of the problems.  For decades our only
insights into quantum gravity came from general relativity and quantum
field theory on spacetime with a fixed background metric.  Now we can
view it from a third angle, that of topological quantum field theory.  
Surely it makes sense to invest some effort in trying to combine
the best aspects of all three theories! 

\begin{figure}[ht]
\vskip 2em
\centerline{\epsfysize=2in\epsfbox{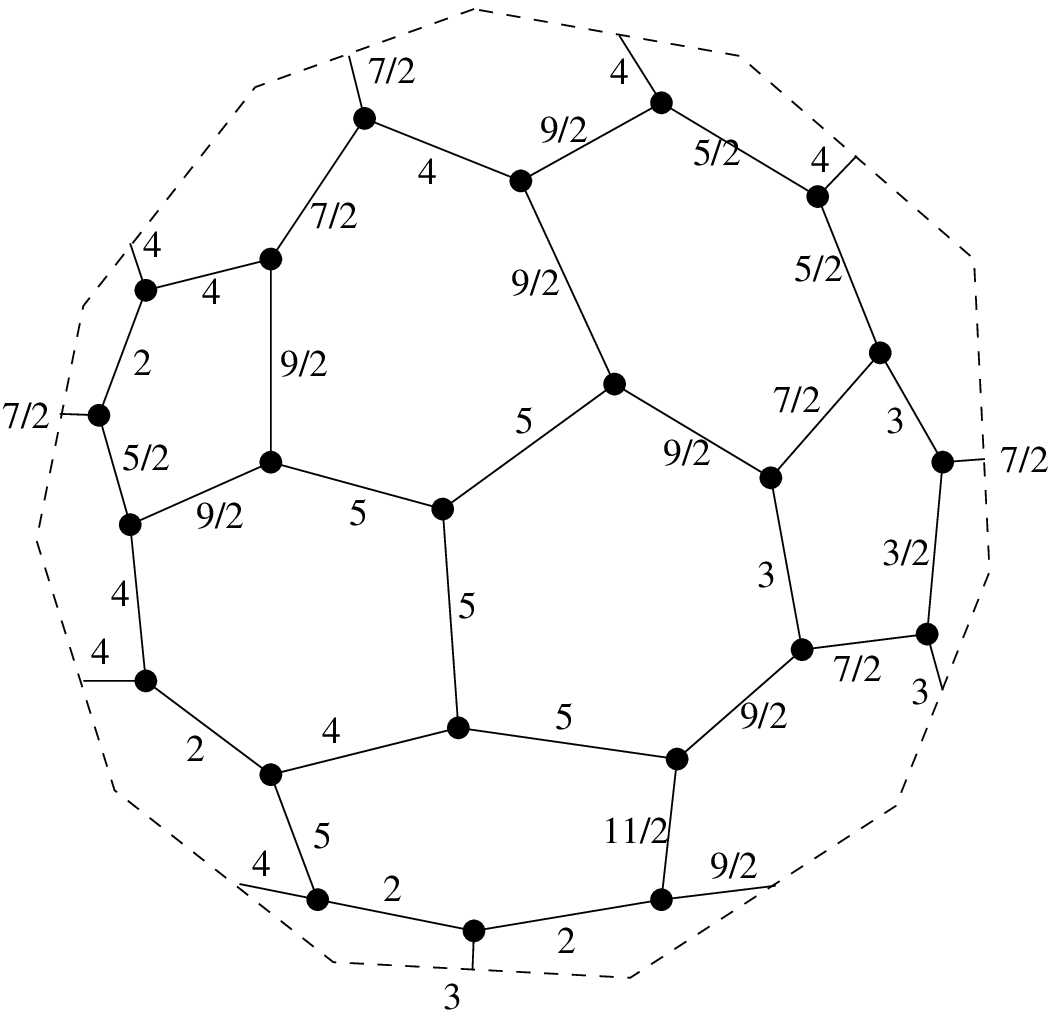}} \medskip
\caption{A spin network}
\label{spin-network}
\end{figure}

And indeed, in the last few years various people have begun to do just
this, largely motivated by tantalizing connections between topological
quantum field theory and loop quantum gravity.   In loop quantum
gravity, the preliminary Hilbert space has a basis given by `spin
networks' --- roughly speaking, graphs with edges labelled by spins
\cite{B,Smolin}.  We now understand quite well how a spin network 
describes a quantum state of the geometry of space.   But spin networks
are also used to describe states in TQFTs, where they arise naturally
from considerations of higher-dimensional algebra.  For example, in
3-dimensional quantum gravity the state shown in Fig.\ 
\ref{triangulated-surface} can also be described using the spin network
shown in Fig.\ \ref{spin-network}.   

Using the relationships between 4-dimensional quantum gravity and
topological quantum field theory, researchers have begun to formulate
theories in which the quantum geometry of spacetime is described using
`spin foams' --- roughly speaking, 2-dimensional structures made of
polygons joined at their edges, with all the polygons being labelled by
spins \cite{B3,BC,FK,Reis,RR}.   The most important part of a spin foam
model is a recipe assigning an amplitude to each spin foam.  Much as
Feynman diagrams in ordinary quantum field theory describe processes by
which one collection of particles evolves into another, spin foams
describe processes by which one spin network evolves into another. 
Indeed, there is a category whose objects are spin networks and whose
morphisms are spin foams!  And like $n\Cob$, this category appears to
arise very naturally from purely $n$-categorical considerations.   

In the most radical approaches, the concepts of `space' and `state' are
completely merged in the notion of `spin network',  and similarly the
concepts of `spacetime' and `process' are merged in the notion of `spin
foam', eliminating the scaffolding of a spacetime manifold entirely.  To
me, at least, this is a very appealing vision.  However, there are a
great many obstacles to overcome before we have a full-fledged theory
of quantum gravity along these lines. Let me mention just a few of the
most pressing.   First there is the problem of developing quantum
theories of Lorentzian rather than Riemannian metrics.  Second, and
closely related, we need to better understand the concept of `causal
structure' in the context of spin foam models.  Only the work of
Markopoulou and Smolin \cite{MS} has addressed this point so far. 
Third, there is the problem of formulating physical questions in these
theories in such a way that divergent sums are eliminated.  And fourth,
there is the problem of developing computational techniques to the point
where we can check whether these theories approximate general relativity
in the limit of large distance scales --- i.e., distances much greater
than the Planck length.  Starting from familiar territory we have
sailed into strange new waters, but only if we circle back to the
physics we know will the journey be complete.  

\subsection*{Acknowledgements}

Conversations and correspondence with many people have helped form my
views on these issues.  I cannot list them all, but I especially want to
thank Abhay Ashtekar, John Barrett, Louis Crane, James Dolan, Louis
Kauffman, Kirill Krasnov, Carlo Rovelli, and Lee Smolin.

\end{document}